\documentclass[aps, prd, notitlepage, secnumarabic, reprint, nofootinbib]{revtex4-1}
\usepackage{amsmath}
\usepackage{bm, braket}
\usepackage{graphicx}
\usepackage[breaklinks=true, colorlinks=true,linkcolor=blue]{hyperref}
\usepackage{CJK}
\usepackage{epsfig}

\begin{document}
\begin{CJK*}{UTF8}{}
\title{Physical renormalization condition for de Sitter QED}

\author{Takahiro Hayashinaka (\CJKfamily{bsmi}林中貴宏)}
\email[Until March 2018, ]{hayashinaka@resceu.s.u-tokyo.ac.jp}
\email[After April 2018, ]{takahiro.hayashinaka@gmail.com}
\affiliation{Research Center for the Early Universe (RESCEU), Graduate School of Science, The University of Tokyo, Bunkyo-ku, Tokyo, 113-0033, Japan}
\affiliation{Department of Physics, Graduate School of Science, The University of Tokyo, Bunkyo-ku, Tokyo, 113-0033, Japan}
\author{She-Sheng Xue (\CJKfamily{gbsn}薛社生)}
\email{xue@icra.it}
\affiliation{ICRANet, Piazzale della Repubblica, 10-65122, Pescara,
Physics Department, Sapienza University of Rome, Piazzale Aldo Moro 5, 00185 Rome, Italy} 
\date{\today}

\begin{abstract}
We considered a new renormalization condition for the vacuum expectation values of the scalar and spinor
currents induced by a homogeneous and constant electric field background in de Sitter spacetime.
Following a semiclassical argument, the condition named maximal subtraction imposes the exponential suppression
on the massive charged particle limit of the renormalized currents.
The maximal subtraction changes the behaviors of the induced currents previously obtained by
the conventional minimal subtraction scheme.
The maximal subtraction is favored for a couple of physically decent predictions
including the identical asymptotic behavior of the scalar and spinor currents,
the removal of the infrared (IR) hyperconductivity from the scalar current, and the finite current for the massless fermion.
\end{abstract}

\maketitle
\end{CJK*}

\section{Introduction}
A quantum electrodynamic (QED) system with a homogeneous and constant energy-density electric background field
in de Sitter spacetime has been investigated by many authors in hopes of cosmological application and
finding new aspects of quantum physics in curved spacetime
\cite{PhysRevD.49.6343,Martin:2007bw,1475-7516-2014-04-009,Cai:2014qba,kobayashi2014schwinger,PhysRevD.93.025004,
Hayashinaka:2016qqn,Hayashinaka:2016dnt,PhysRevD.94.104011,Bavarsad:2017oyv}.
Despite the disregard of realistic features in the cosmological context, the system provides a theoretical probe into
the problems of quantum fields in the curved spacetime.

The two nontrivial background fields semiclassically cause the quantum-mechanical particle productions.
The particle production from the electric field, known as Schwinger effect~\cite{schwinger1951gauge},
is one of the nonperturbative QED effects as the number density of the created particles is proportional to the factor $\exp(-\pi m^2/eE)$
(where $m$, $e$, and $E$ are the particle mass, the coupling, and the strength of the homogeneous background electric field, respectively).
Besides, the gravitational field also causes particle creation.
The number density of the created particles in de Sitter spacetime with the Hubble parameter $H$
is proportional to $(\exp(2\pi m/H)\pm1)^{-1}$ ($-$ for boson, $+$ for fermion).
Both approach to $\exp(-2\pi m/H)$ when the boson and fermion are heavy, i.e., $m/H\gg1$.

The vacuum expectation value (VEV) of the electric current $\braket{J^\mu}$ induced by the electric and gravitational fields
is a quantity with firm theoretical ground since it appears on the right-hand side of the equation of motion for the gauge field.
Hence, the induced current describes the backreaction from the charged particles to the background electric field.
During the previous research of the induced currents in de Sitter spacetime, a couple of curious features were discovered.
In the case of scalar QED with small mass parameter $m$, a phenomenon called infrared (IR) hyperconductivity, i.e., the strong enhancement of
the scalar current for the weaker electric field was found first in the $1+1$ dimensional problem~\cite{1475-7516-2014-04-009},
and then in $1+2$ dimension~\cite{PhysRevD.94.104011} and $1+3$ dimension~\cite{kobayashi2014schwinger,Hayashinaka:2016dnt}.
Besides, in the studies in $1+3$ dimensional de Sitter spacetime, a negative current which flows in the direction opposite to
the electric field was found for the scalar case~\cite{kobayashi2014schwinger,Hayashinaka:2016dnt} and the fermionic case~\cite{Hayashinaka:2016qqn}.
This is surprising since QED is a screening theory though the negative current indicates the anti-screening of the electric field.
Also, the terms which are not suppressed by the exponential factors, $\exp(-\pi m^2/eE)$ or $\exp(-2\pi m/H)$,
was found in the massive field limit $m/H\gg1$ of the induced current.
These suppression factors are naturally expected from the semiclassical approximation.
Therefore, it is rational to be doubtful of the validity of the results in $1+3$ dimension.

The skepticism about the $1+3$ dimensional results raises a concern with the renormalization scheme employed in the previous calculations.
The adiabatic subtraction scheme~\cite{PhysRevD.9.341} with the gauge-breaking momentum cutoff was adopted in~\cite{kobayashi2014schwinger,Hayashinaka:2016qqn}.
Its implementation is simple, but the gauge invariance is apparently violated.
In~\cite{Hayashinaka:2016dnt}, a gauge-invariant and covariant calculation of the scalar induced current based on
the point-splitting technique was done, and the correspondence between the two techniques was confirmed.
An aspect vital to the agreement between the two regularization schemes was that the subtraction from the VEV remained minimal.
In other words, imposing minimal subtraction assured the consistency of the different renormalization schemes.

We propose an alternative to the renormalization condition in the present article.
The new condition, which we call maximal subtraction, requires that the asymptotic behavior
of the renormalized VEV should obey the semiclassical approximation.
That is, we impose the exponential damping of the induced currents by the factors $\exp(-\pi m^2/eE)$ or $\exp(-2\pi m/H)$
provided the parameters satisfy the semiclassical condition $(eE/H^2)^2+(m/H)^2\gg1$.

In the next section \ref{review_minimal_subtraction}, we introduce the precise setup and notations.
Then, we review the calculation procedure of the previous results while clarifying the renormalization condition employed.
In section \ref{maximal_subtraction_details}, the maximal subtraction scheme is applied, and detailed analysis is shown.
The comparison between the maximal subtraction and the minimal subtraction is made in section \ref{results_max_min_subtraction}.
In addition, we provide the phase diagram which exhibits parameter regions of the positive and negative current values.
Lastly, in section \ref{conclusion_max_min_subtraction}, we conclude the present article with remarks about
the physical origin of the negative current.

\section{Review of the previous results}\label{review_minimal_subtraction}
\subsection{Induced current}
The system we consider consists of a charged scalar $\phi$ or a Dirac field $\psi$, a gravitational background
whose metric is given by $g_{\mu\nu}=a^2(\eta)\mathrm{diag}(-1,1,1,1)$, and an electric background field
$A_\mu^{\mathrm{bg}} = (0,0,0,A_z(\eta))$ in $z$-direction.
We will work in de Sitter spacetime where the scale factor is given by $a=\mathrm e^{Ht}$ with
the constant Hubble parameter $H$.
The cosmic time $t$ and the conformal time $\eta$ are connected by $\mathrm dt = a\mathrm d\eta$. 

We consider the scalar (boson) and spinor (fermion) QED actions $S_{b,f}$ separately.
In terms of the canonical matter field variables $\chi\equiv a\phi$ and $\xi\equiv a^{3/2}\psi$,
these actions are given by
\begin{equation}\begin{split}\label{action_scalarQED_canonical}
S_b = &\int \mathrm d\eta \mathrm d^3x
\bigl\{ 
\mathcal L_{g}
- \eta^{\mu\nu}\chi^\dagger_{,\mu}\chi_{,\nu} 
+ ie\eta^{\mu\nu}A_\mu\chi^\dagger\overleftrightarrow{\partial_\nu}\chi\\
&\quad -(\eta^{\mu\nu}e^2A_\mu A_\nu+(m_b^2-2H^2)a^2)\chi^\dagger\chi \bigl\},
\end{split}\end{equation}
and
\begin{equation}\label{action_spinorQED_canonical}
S_f = \int \mathrm d\eta \mathrm d^3x
\left\{ 
\mathcal L_{g}
+ \bar\xi \left( i \gamma^a\partial_a - eA_a\gamma^a - m_f a \right)\xi\right\},
\end{equation}
where $\eta^{\mu\nu}$ denotes the Minkowski metric with sign convention $(-,+,+,+)$,
$\mathcal L_g = -1/4\eta^{\mu\alpha}\eta^{\nu\beta}F_{\mu\nu}F_{\alpha\beta}$ is the gauge kinetic term,
$F_{\mu\nu} = A_{\nu,\mu}-A_{\mu,\nu}$ is the field strength,
and the gamma matrix $\gamma^a$ satisfies $\{\gamma^a,\gamma^b\}=-2\eta^{ab}$.
Note that $A_a=\delta^\mu_a A_\mu$.
The Dirac conjugate is defined as $\bar\xi = \xi^\dagger\gamma^0$.

The equations of motion for the charged fields $\chi$ (Klein-Gordon equation) and $\xi$ (Dirac equation) are
easily obtained from the actions.
Hereafter, we set $m_b=m_f=m$ to make the notation simpler.
The equations were solved analytically in~\cite{kobayashi2014schwinger,Hayashinaka:2016dnt,Hayashinaka:2016qqn}
with the canonical quantization condition
$[\hat\chi(\eta,\bm x), \frac{\mathrm d}{\mathrm d\eta}\hat\chi^{\dagger}(\eta,\bm y)]
=\{\hat\xi(\eta,\bm x),i\hat\xi^\dagger(\eta,\bm y)\}=i\delta^{(3)}(\bm x-\bm y)$ 
and the background assumption
\begin{equation}\label{backgroundAz}
A_\mu=A_\mu^{\mathrm{bg}}\equiv \left(0,0,0,-\frac{E}{H}(a-1)\right),
\end{equation}
which yields the homogeneous and constant energy density electric field in $z$-direction.
The electric field strength is characterized by the parameter $E$.
Besides, the equation of motion for the gauge field (Maxwell's equation)
is sourced by the bosonic current
\begin{equation}
J_b^\mu = ie\eta^{\mu\nu}(\chi^\dagger\chi_{,\nu} - \chi_{,\nu}^\dagger\chi+2ieA_\nu\chi^\dagger\chi),
\end{equation}
or the fermionic current
\begin{equation}
J_f^\mu = -e \bar\xi \delta_{a}^{\mu}\gamma^a \xi.
\end{equation}
The backreaction to the gauge field defined as $A_\mu^{\mathrm{br}}=A_\mu-A_\mu^{\mathrm{bg}}$ is then described
by the renormalized VEV of the current operators $\braket{\hat J_{b,f}^\mu}_{\mathrm{ren}}$
induced by the background gauge field, so we have
\begin{equation}\label{backreaction_eom_renormalized}
F^{\mu\nu}_{\;\;,\nu}\bigl|^{\mathrm{br}} \equiv (A^{\mathrm{br}})^{\nu,\mu}_{\;\;,\nu} - (A^{\mathrm{br}})^{\mu,\nu}_{\;\;,\nu}
= \braket{\hat J_{b,f}^\mu}_{\mathrm{ren}}.
\end{equation}
The unrenormalized version of Eq.~\eqref{backreaction_eom_renormalized},
$F^{\mu\nu}_{\;\;,\nu}=\braket{\hat J_{b,f}^\mu}$, should be examined
to renormalize the divergence appears in the raw VEV of the induced current.
According to the gauge-invariant calculation performed in~\cite{Hayashinaka:2016dnt}, the divergent part of the VEV
$\braket{\hat J_{b,f}^\mu}_{\mathrm{div}}$ has the same functional dependence as the partial derivative of the
background field strength $F^{\mu\nu}_{\;\;,\nu}$, and can be absorbed into the conventional renormalization
of the gauge field and the charge involving a divergent coefficient $C$ which takes the following form
\begin{equation}
A_{\mathrm R \, \mu} = C A_{\mu}, \quad e_{\mathrm R} = C^{-1} e.
\end{equation}
The explicit form of the renormalization coefficient $C$ is then fixed by the analytical form of $\braket{\hat J_{b,f}^\mu}$.
For example, $z$-component of the raw VEV of the bosonic current is given by
\begin{equation}\label{bosonic_current_div}
\begin{split}
\braket{\hat J_b^z} = \lim_{\epsilon\to0}\frac{e(aH)^3}{4\pi^2}\biggl[&-\frac{L}{3}(\log\epsilon+\log H+\frac{3}{2}+\gamma_E)\\
&+\mathcal O(\epsilon^0) + \mathcal O(\epsilon^1)\biggl]
\end{split}
\end{equation}
where the parameter $\epsilon$ controls the covariant point separation.
See Eq.~(2.15) in~\cite{Hayashinaka:2016dnt} for the full analytical form of the $\mathcal O(\epsilon^0)$ term.
Note that the $\mathcal O(\epsilon^0)$ part is a function of the two dimensionless parameters defined as
\begin{equation}\label{def_L_M}
L\equiv \frac{eE}{H^2}, M\equiv\frac{m}{H}.
\end{equation}
Other components are zero because of the symmetry of the system.
The renormalization coefficient then takes the following form
\begin{equation}\label{renC}
C^2 = 1 - \frac{e^2}{24\pi^2}\left( \log\epsilon + \alpha \right),
\end{equation}
where $\alpha$ denotes the finite part.
Thus, the subtraction of terms proportional to $e (aH)^3 L$ in Eq.~\eqref{bosonic_current_div} can 
be done arbitrarily by fixing $\alpha$.
Ideally, the renormalization condition which determines the finite term is given by experiments or observations.
However, we do not have sufficient experimental knowledge about quantum effects in de Sitter spacetime.
Thus, we can make theoretical predictions only with some physical assumptions.


\subsection{Minimal subtraction}
A physical condition is required to determine the finite part of the subtraction.
We adopt a necessary condition that the renormalized current must vanish for extremely heavy particles ($m^2\gg E,\,H^2$),
so we have $\braket{\hat J_{b}^\mu}_{\mathrm{ren}} \to 0$ for $m\to\infty$.
This condition partially determines the finite part $\alpha$ of Eq.~\eqref{renC} as
$\alpha = \log m + \gamma_E + \frac{3}{2} + \tilde\alpha$.
Here $\tilde\alpha$ must vanish for $m\to\infty$.

The minimal subtraction scheme requires $\tilde\alpha$ should vanish,
which results in the renormalized bosonic current $\braket{\hat J_{b}^\mu}_{\mathrm{ren}}$
where the first term in the large parenthesis $[\cdots]$ of Eq.~\eqref{bosonic_current_div}
is replaced by $(L/3)\log M$.

It was shown, in~\cite{Hayashinaka:2016dnt}, that the result $\braket{\hat J_{b}^\mu}_{\mathrm{ren}}$
obtained by the minimal subtraction scheme coincides with the one obtained by
the minimal-order adiabatic subtraction shown in~\cite{kobayashi2014schwinger},
which naturally follows a general proof of the coincidence between the point-splitting renormalization
and the adiabatic subtraction with the minimal order~\cite{PhysRevD.9.341}.
For the fermionic induced current, the evaluation of the VEV $\braket{\hat J_{f}^\mu}_{\mathrm{ren}}$
was only done by the adiabatic subtraction, which is given by Eq.~(3.12) in~\cite{Hayashinaka:2016qqn}.
Note that the agreement of the point-splitting renormalization and adiabatic subtraction is
also proven for the general fermionic case in~\cite{PhysRevD.91.064031}.

The properties and consequences of the results with minimal subtraction was deeply investigated
in~\cite{Hayashinaka:2016qqn,Hayashinaka:2016dnt} where two enigmatic behaviors were revealed.
One is the negative current, that is, the anomalous charge transportation in the direction
opposite to the background electric field.
The negative current indicates the anti-screening and the instability of the system we consider.
A possible physical explanation of this effect will be the subject of the future work.
The other is the unexpected asymptotic behavior of the currents in the massive particle limit,
which leads to the reconsideration of the minimal subtraction.

\subsection{Massive limit}
Hereafter, we will focus on the dimensionless currents defined as
$\mathcal J_{b,f} (L, M) \equiv \Braket{\hat J^z_{b,f}}_{\mathrm{ren}}/(ea^3H^3)$,
which are depending only on the dimensionless electric field strength and mass defined in Eq.~\eqref{def_L_M}.
Furthermore, we introduce the dimensionless conductivities $\sigma_{b,f}(M) \equiv \mathcal J_{b,f}/L|_{L\to0}$
at zero electric field state.
The analytical form of the bosonic conductivity $\sigma_b$ is given by
\begin{equation}\begin{split}\label{sigma_boson_minimal}
\sigma_b(M) &= \frac{1}{12\pi^2} 
\biggl\{ \frac{8\pi}{3}\frac{\mu_0(\mu_0^2-1)}{\sin(2\pi\mu_0)}+\log M \\
& -\frac{1}{2}\left[\psi\left(\frac{1}{2}+\mu_0\right)+\psi\left(\frac{1}{2}-\mu_0\right)\right]\biggl\},
\end{split}\end{equation}
where $\mu_0\equiv\sqrt{9/4-M^2}$, and $\psi(z)$ denotes the digamma function.
The fermionic conductivity $\sigma_f$ is given by
\begin{equation}\label{sigma_fermion_minimal}
\sigma_f(M) = \frac{1}{3\pi^2} \left[ \log M - \frac{\pi M(4M^2+1)}{3\sinh(2\pi M)} - \Re \psi(iM) \right].
\end{equation}

The dimensionless conductivities have the following asymptotic behavior for $M\to\infty$
\begin{equation}\label{sigma_bf_minimal_massive}
\sigma_{b,f}(M) =
\mathrm e^{-2\pi M}\sum_{i=0}^{\infty} c_i M^{3-2i} + \sum_{i=0}^{\infty} d_i M^{-2-2i},
\end{equation}
where the leading coefficients $(c_1,d_1)$ are given by $(-4/(9\pi),7/(72\pi^2))$ for bosons
and $(-8/(9\pi),-1/(36\pi^2))$ for fermions.
Because the standard Bogoliubov calculation providing the number densities of the scalar and Dirac particles in de Sitter spacetime,
$n\sim H^4 (\mathrm e^{2\pi M} \pm 1)^{-1}$, is valid in the semiclassical limit $L^2+M^2\gg1$,
the induced currents and conductivities must inherit the overall factor
$\mathrm e^{-2\pi M}$ in the massive field limit.
Nevertheless, the latter terms ($d_i$-terms) which are not protected by the exponential factor
is also seen in the asymptotic expansion Eq.~\eqref{sigma_bf_minimal_massive}.


\section{New renormalization condition}\label{maximal_subtraction_details}
\subsection{Maximal subtraction}
\begin{figure}[hb]
 \begin{center}
  \includegraphics[width=0.5\textwidth]{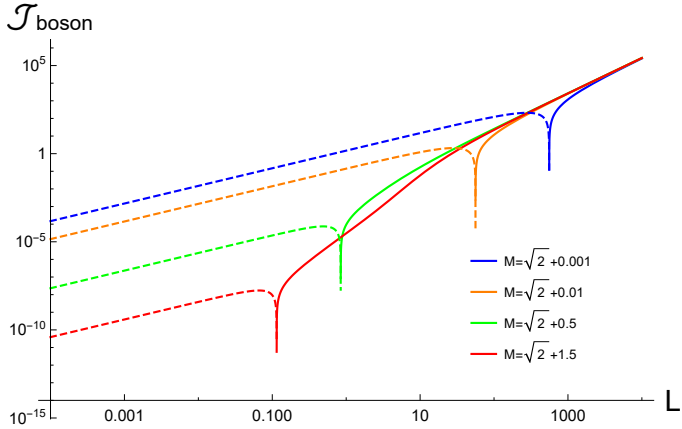}
 \end{center}
 \caption{The maximally-subtracted bosonic current.
 The IR hyperconductivity observed in the minimal subtraction is removed.
 Instead, the enhancement of the negative current in the conformally
 massless limit $M\to\sqrt 2$ is seen.}
\label{fig:scalar_current_max}
\end{figure}

\begin{figure}[htb]
 \begin{center}
  \includegraphics[width=0.5\textwidth]{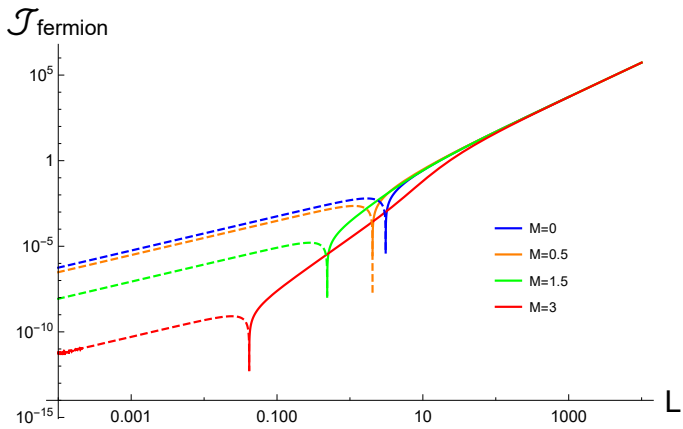}
 \end{center}
 \caption{The maximally-subtracted fermionic current. 
 The finite behavior of the massless fermion is seen.
 The behavior of the massive fermion is identical to the massive boson in the maximal subtraction scheme.}
\label{fig:spinor_current_max}
\end{figure}

\begin{figure}[htb]
 \begin{center}
  \includegraphics[width=0.5\textwidth]{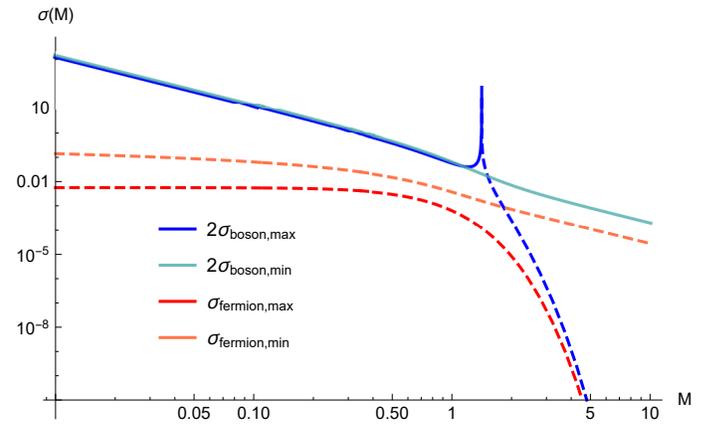}
 \end{center}
 \caption{The conductivities $\sigma_{(b,f),(\mathrm{min},\mathrm{max})}$ given in
 Eq.~\eqref{sigma_bf_max} (blue for maximally-subtracted boson, and red for maximally-subtracted fermion),
 Eq.~\eqref{sigma_boson_minimal} (light green for minimally-subtracted boson), and
 Eq.~\eqref{sigma_fermion_minimal} (orange for minimally-subtracted fermion).
 Factor $2$ is introduced for the bosonic conductivity to compensate the difference in the spin degrees of freedom.} 
\label{fig:sigma_bf_max_min}
\end{figure}

The emergence of the $d_i$-terms is not only unexpected but also inexplicable.
Thus, we perceive the $d_i$-terms possibly being renormalization artifacts which should be removed.
This assumption is rephrased as imposing the exponential mass suppression
on the renormalized currents $\mathcal J_{b,f}$.
This is implemented by an extra subtraction of the terms from the minimally-subtracted results
$\mathcal J_{b,f,\mathrm{min}}$ shown in the previous section.
We name this maximal subtraction.
As the logarithmic and digamma functions in Eq.~\eqref{sigma_boson_minimal} and Eq.~\eqref{sigma_fermion_minimal}
are responsible for the $d_i$-terms,
the maximal subtraction of the bosonic current is implemented as follows
\begin{equation}\begin{split}\label{current_max_boson}
\mathcal J_{b,\mathrm{max}} = &\mathcal J_{b,\mathrm{min}}- \frac{L}{12\pi^2} \biggl\{ \log M \\
& - \frac{1}{2}\left[\psi\left(\frac{1}{2}+\mu_0\right)+\psi\left(\frac{1}{2}-\mu_0\right)\right]\biggl\},
\end{split}\end{equation}
while the maximal subtraction for the fermionic current is given by
\begin{equation}\label{current_max_fermion}
\mathcal J_{f,\mathrm{max}} = \mathcal J_{f,\mathrm{min}} -\frac{L}{3\pi^2} \left[ \log M - \Re \psi(iM)\right].
\end{equation}
The resulting currents as functions of the electric field strength $L$ is shown
in Fig.~\ref{fig:scalar_current_max} (boson) and Fig.~\ref{fig:spinor_current_max} (fermion).
In the Fig.~\ref{fig:scalar_current_max}, Fig.~\ref{fig:spinor_current_max}, and also in Fig.~\ref{fig:sigma_bf_max_min},
the solid (dashed) lines indicates the positive (negative) values.
See the minimally-subtracted results Fig.1 in~\cite{Hayashinaka:2016qqn} for comparison.

The bosonic and fermionic conductivities are also obtained as
\begin{equation}\begin{split}\label{sigma_bf_max}
\sigma_{b,\mathrm{max}}(M) &= \frac{2}{9\pi}\frac{\mu_0(\mu_0^2-1)}{\sin(2\pi\mu_0)},\\
\sigma_{f,\mathrm{max}}(M) &= -\frac{M(4M^2+1)}{9\pi\sinh(2\pi M)}.
\end{split}\end{equation}
The behaviors of the minimally- and maximally-subtracted conductivities are shown in Fig.~\ref{fig:sigma_bf_max_min}.

The maximally-subtracted bosonic conductivity has a discontinuity at $M=\sqrt 2$ which corresponds to
the conformal coupling $\xi R\phi^2$ with the parameters being $\xi=1/6$ and $R=12H^2$ in de Sitter spacetime.
Thus, this is conformally equivalent to the case of a massless scalar field in Minkowski spacetime.
For this reason, the discontinuity and divergence at $M=\sqrt 2$ are reasonable.

\subsection{Physical motivations of the maximal subtraction}
One might discard the possibility of the maximal subtraction for the reason that
it brings about the divergence to the bosonic conductivity.
Although the aversion to the non-analyticity may be reasonable,
it is also true that there is no excuse to accept the minimal subtraction scheme without due deliberation.
On the contrary, there are a couple of facts which support the maximal subtraction against the minimal subtraction.

The primary motivation is the appearance of the $d_i$-terms in the massive limit of the currents and conductivities
shown in Eq.~\eqref{sigma_bf_minimal_massive} as mentioned.
From the semiclassical viewpoint, one expects of the exponential factor $\exp(-2\pi M)$ in
the quantities proportional to the particle number density.
The maximal subtraction can entirely fix the form of the subtraction term
and immediately remove the counterintuitive terms from the VEV of the currents.

The second motivation is the discrepancy between the bosonic and fermionic currents with the minimal subtraction in the massive limit despite the identical behavior for the massive boson and fermion.
The particle picture is validated in the semiclassical limit which is conditioned as $L^2+M^2\gg1$.
In the case of the strong field $L\gg1$, the bosonic and fermionic currents behave similarly
except for the factor $2$ originates from the spin degrees of freedom and take the limiting form
\begin{equation}\label{strong_lim_current}
\mathcal{J}_{f,\mathrm{min}} \simeq 2\mathcal{J}_{b,\mathrm{min}}
\simeq \frac{L^2}{6\pi^3}\mathrm e^{-\frac{\pi M^2}{L}} = \frac{(eE)^2}{6\pi^3H^{4}} \mathrm e^{-\frac{\pi m^2}{eE}}, 
\end{equation}
where we recover the factor $\mathrm e^{-\pi m^2/eE}$ of the Schwinger mechanism in the flat spacetime.
However, if we consider the massive fields $M\gg1$ while keeping the electric field relatively weaker $L\lesssim1$,
then the minimally-subtracted bosonic and fermionic currents behave quite differently~\footnote{
This discrepancy is not only quantitative but also qualitative,
that is, the bosonic conductivity $\sigma_{b,\mathrm{min}}$ is positive though the fermionic conductivity $\sigma_{f,\mathrm{min}}$ is negative
This is unfavored also because the bosonic current becomes positive in this limit despite the intuitive
explication for the negativity of the current, or the anti-screening effect we recently obtained, which is independent of the spin of charged particles.
This physical interpretation of the origin of the negative current is shown
in TH's Ph.D. thesis~\cite{phdthesis_Hayashinaka_2018_UTokyo} and will be published in an article.
See also the conclusion section.
}.
The maximal subtraction cures the disagreement and yields the identical limiting behavior for $M\gg1$
\begin{equation}
\mathcal{J}_{f,\mathrm{max}} \simeq 2\mathcal{J}_{b,\mathrm{max}}
\simeq - \frac{8}{9\pi}M^3\mathrm e^{-2 \pi M} L.
\end{equation}
Since the maximal subtraction does not affect the strong electric field behavior Eq.~\eqref{strong_lim_current},
the new renormalization condition realizes the boson-fermion agreement
in all the semiclassical regime $L^2+M^2\gg1$.

\section{Results}\label{results_max_min_subtraction}
There are four striking features of the maximally-subtracted induced currents plotted in
Fig.~\ref{fig:scalar_current_max} and Fig.~\ref{fig:spinor_current_max}.

The first point is the exponential damping of the currents for $M\gg1$.
This is trivial since we have imposed it as the renormalization condition.
The second point is the boson/fermion agreement in all the semiclassical regime as we mentioned
in the last part of the Sec.~\ref{maximal_subtraction_details}.
The third point is the removal of the IR hyperconductivity from the bosonic current.
Instead, we see the enhancement of the bosonic conductivity in the conformally massless boson limit $M\to\sqrt{2}$,
i.e., $\sigma_{b,\mathrm{max}}\propto(M-\sqrt{2})^{-1}$.
The final point is the finite behavior of the massless fermionic current similar to the $2$-dimensional case \cite{PhysRevD.93.025004}.

\begin{figure}[htb]
 \begin{center}
  \includegraphics[width=0.5\textwidth]{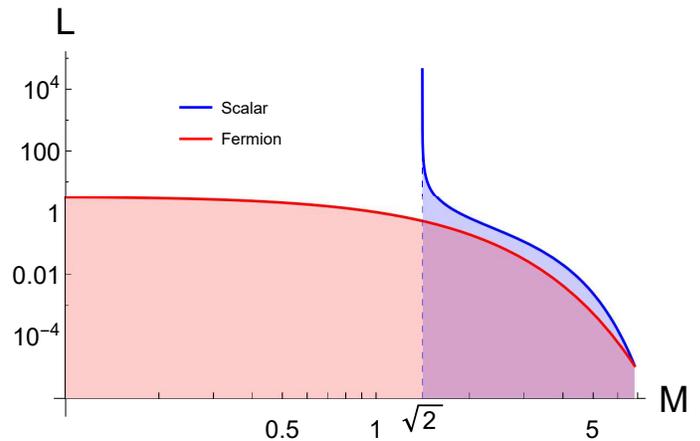}
 \end{center}
 \caption{The phase diagram for the scalar (blue) and spinor (red) QED in de Sitter spacetime obtained by the maximal subtraction.
 The shaded regions correspond to the negative current phase, or the anti-screening phase,
 while the regions above the curves correspond to the positive current phase with standard screening effects.
 See also the minimal subtraction counterparts, Fig.4 in~\protect{\cite{Hayashinaka:2016dnt}}
 and Fig.2 in~\protect{\cite{Hayashinaka:2016qqn}}.
 }
\label{fig:phase_max}
\end{figure}

The change in the renormalization condition affects the IR behavior of the induced current. 
Determination of the electric field strength $L_\ast$ which causes no backreaction, i.e., $\mathcal J_{b,f}(L_\ast,M)=0$,
is of particular interest because the critical value $L_\ast(M)$ corresponds to a stable point or
a saddle point of the current-electric field system described by the backreaction equation \eqref{backreaction_eom_renormalized}.
This critical field strength is indicated by the lines in Fig.~\ref{fig:phase_max}.
The positive current causes negative backreaction, so it reduces the background electric field.
This illustrates the general screening nature of (quantum)electrodynamics.
Conversely, the negative current causes the anti-screening effect, and it will increase
the background electric field strength.
The shaded regions below the lines in Fig.~\ref{fig:phase_max} correspond to the anti-screening phase.
Therefore, the two lines in Fig.~\ref{fig:phase_max} indicate the stable points of the system.
Adding to the existence of the nontrivial stable points $L_\ast$,
$L=0$ is the trivial zero of the induced currents, i.e., $\mathcal J_{b,f}(L=0,M)=0$
(see the full analytical forms of the induced currents shown in~\cite{kobayashi2014schwinger,Hayashinaka:2016dnt,Hayashinaka:2016qqn}).
So, the state of no electric field corresponds to the saddle point of the system, and cannot be the equilibrium state.
The existence of the nonzero stable values $L_\ast(M)\neq0$ implies the spontaneous generation of
the large-scale electric field in de Sitter spacetime.
This possibility may open a new scenario of primordial generation of the electromagnetic field
in the inflationary universe by the anti-screening effect of QED.

\section{Conclusion}\label{conclusion_max_min_subtraction}
We studied a possibility of an alternative renormalization condition for the VEV of the scalar and spinor currents
induced by the electric background field Eq.~\eqref{backgroundAz} in de Sitter spacetime.
The renormalization condition we proposed is the maximal subtraction which requires that the asymptotic behavior of the VEV
should match up to that of the semiclassical approximation.
Focusing on the massive limit of the minimally-subtracted electric conductivities in Eq.~\eqref{sigma_bf_minimal_massive},
we specified the terms which should be further subtracted.
The maximally-subtracted results are given by Eq.~\eqref{current_max_boson} for the scalar current
and Eq.~\eqref{current_max_fermion} for the Dirac current.
These results are exhibited in Fig.\ref{fig:scalar_current_max} and Fig.\ref{fig:spinor_current_max}, respectively.
Predictably, the resulted conductivities are exponentially suppressed in the semiclassical limit
$L^2+M^2\gg1$ as shown in Fig.\ref{fig:sigma_bf_max_min},
which also shows the boson-fermion agreement and the negativity of the maximally subtracted conductivities.
The phase diagram based on the linear response analysis is also exhibited in Fig.~\ref{fig:phase_max}.

The physical explanation of the negative induced current for the massive charged particle has not been given
in the present article, but it is explained by analogy with the Hawking radiation in~\cite{phdthesis_Hayashinaka_2018_UTokyo}.
In short, the de Sitter horizon can separate the particles from anti-particles assisted by the electric field
and generate a charge density distribution on the horizon.
As the generated charge density distribution creates the polarization in the direction opposite to
the electric field, it can cause the anti-screening effect.
This action of the horizon can be compared to that of the thermodynamical Maxwell's demon who has
access to the microscopic information of particles to sift them from fluctuations according to the particle attributes.
A microscale battery which realizes Maxwell's demon who extracts the electric energy from the thermal motion of gas molecules has been developed recently~\cite{Chida2017}.

The properties of the maximal subtraction indicate the correctness of our assumption.
In~\cite{Urakawa:2009xaa}, it was discussed that cosmologically observable quantities are
insensitive to the precise renormalization condition.
In contrast, we have revealed that a physical process during the inflation can be
affected qualitatively as well as quantitatively by choice of the renormalization condition.
The preference of the maximal subtraction over the minimal-order adiabatic subtraction
which has been employed in many pieces of literature, e.g.,~\cite{kobayashi2014schwinger, PhysRevD.9.341, PhysRevD.36.2963,
PhysRevD.89.044030, PhysRevD.91.124075, birrell1978application, PhysRevD.80.065024,PhysRevD.76.103528, AALambitThesis2017},
questions its validity.
Remarkably, some authors reported the unphysical consequences of the minimal subtraction scheme such as
the negative value of the renormalized primordial power spectrum~\cite{PhysRevD.76.103528, AALambitThesis2017}.

There is no doubt at the end of the day that the correct renormalization condition is obtained only experimentally or observationally.
On the other hand, it seemingly makes sense that matching
the asymptotic behavior of a physical quantity to its semiclassical estimation.

\begin{acknowledgments}
This work was supported by JSPS KAKENHI, Grant-in-Aid for JSPS Fellows 15J09390 (TH). 
\end{acknowledgments} 

\bibliography{references}

\end{document}